# Transformation Ray Method: Controlling High Frequency Elastic Waves

Zheng Chang[1], Jin Hu[2], Xiaoning Liu[1] and Gengkai Hu[1,*]

1   School of Aerospace Engineering, Beijing Institute of Technology, Beijing 100081, People's Republic of China.

2   School of Information and electronics, Beijing Institute of Technology, Beijing 100081, People's Republic of China.

**ABSTRACT**：A transformation method based on elastic ray theory is proposed to control high frequency elastic waves. We show that ray path can be controlled in an exact manner, however energy distribution along the ray is only approximately controlled. A numerical example of an elastic rotator is provided to illustrate the method and to access the approximation. The proposed theory may be found potential applications in seismic wave protection and structure health monitoring.

---

[*] hugeng@bit.edu.cn

Transformation method proposed by Greenleaf et al. [1], Pendry et al. [2] and Leonhardt [3], provides a direct solution for finding spatial material distribution in order to control electromagnetic wave in a desired way. The basic principle lies in form-invariance of Maxwell equations under a general spatial transformation, so an effect of topological change can be mimicked by a material spatial distribution. The same idea is also applied to acoustic wave since Helmholtz equation is form invariant [4-7]. However for elastic waves in solid materials, Milton et al. show that Navier's equation is transformed to Willis' one instead under a general spatial transformation [8], therefore the techniques developed for electromagnetic or acoustic waves fail to be applied directly for solid elastic waves. Several works have been conducted to control elastic wave by placing spatially the material, for example, Brun et al. [9] proposed a cylindrical cloak for in-plane elastic wave with an asymmetric elasticity tensor. Zhou et al. [10] suggested to use impedance matched method to construct a cloak for shielding elastic wave in long wavelength limit. Amirkhizi et al. [11] use the anisotropy of material to guide elastic wave in preferred directions. Hu et al. [12] proposed an approximate transformation method for elastic wave based on local affine assumption for mapping. Up to date, the theory on controlling elastic waves is far from mature. In this letter, we will examine the transformation method in context of elastic ray theory, which is valid for high frequency elastic waves. The transformation relations for material parameters will be derived from the eikonal equation, and they will be shown to verify only approximately the transport equation. The approximation will be also addressed by comparing with full wave simulation.

We start from the general elastodynamics equation (Navier's equation):

$$(C_{ijkl}u_{k,l})_{,j} = -\rho_{ij}\ddot{u}_j, \qquad (1)$$

where **u** is the displacement vector and **C** the rank-four elastic tensor. The mass density $\rho$ is a second-order tensor to include possible metamaterials. For an inhomogeneous medium, usually Navier's equation cannot be solved with several independent wave modes. However in a smoothly inhomogeneous media, a high-frequency elastic wave can be approximately separated into P and S waves, called respectively quasi-compressional (qP) wave and quasi-shear (qS1 and qS2) waves. In this case, to solve (1), a time-harmonic solution is represented in form of ray series [13]:

$$u_i(x_j, t) = \left[\sum_{n=0}^{\infty} \frac{U_i^{(n)}(x_j)}{(-i\omega)^n}\right] \exp(-i\omega(t - T(x_j))), \qquad (2)$$

where $\mathbf{U}^{(n)}$ is the amplitude vector of the n-th order and $T$ is a scalar function, called travel time (eikonal). In the following discussion, we will follow the zeroth-order approximation of ray series, as in the reference [13].

A harmonic plane wave solution of displacement vector is assumed to be of the following form:

$$u_i(x_j, t) = U_i(x_j) \exp(-i\omega(t - T(x_j))). \qquad (3)$$

Inserting (3) into (1) yields

$$(C_{ijkl} U_{k,l})_{,j} + i\omega[C_{ijkl} U_{k,l} T_{,j} + (C_{ijkl} U_k T_{,l})_{,j}] - \omega^2[(C_{ijkl} U_k T_{,l} T_{,j}) - \rho_{ij} U_j] = 0. \qquad (4)$$

Equation (4) should be satisfied for any frequency, so the coefficients with $\omega^n$ ($n = 0, 1, 2$) must vanish, we get:

$$(C_{ijkl} U_k T_{,l} T_{,j}) - \rho_{ij} U_j = 0, \qquad (5a)$$

$$C_{ijkl} U_{k,l} T_{,j} + (C_{ijkl} U_k T_{,l})_{,j} = 0, \qquad (5b)$$

$$(C_{ijkl} U_{k,l})_{,j} = 0. \qquad (5c)$$

For high frequency elastic waves ($\omega \gg \omega^0 = 1$), the first term in (4) (with $\omega^0$) can be neglected compared to the second (with $\omega^1$) and third (with $\omega^2$) terms, therefore it can be safely dropped [13]. So, the governing equations for elastic ray theory consist of (5a) and (5b), they are complete with initial and boundary conditions. (5a) is the governing equation for the path of elastic ray, called eikonal equation; while (5b) monitors the energy transfer along the ray, called transport equation. In the following, we will examine the form invariance of the governing equations for elastic ray under a general mapping.

Consider a general mapping $x'_{i'} = x'_{i'}(x_j)$, (5a) is transformed into $x'$ coordinates as

$$(\beta_j^{j'} \beta_l^{l'} C_{ijkl} U_k T_{,l'} T_{,j'}) - \rho_{ij} U_j = 0, \qquad (6)$$

where $\beta_i^{i'} = \partial x'_{i'}/\partial x_i$ is the component of Jacobin matrix of the coordinate transformation, $T$ is a scalar independent on the coordinates. Similarly, expressing (5b) in $x'$ coordinates, and introducing the Jacobian $J$ of the coordinate transformation, and with help of the relation $(J^{-1} \beta_{i'}^{i})_{,i} = 0$ [5, 14], we get

$$J^{-1} \beta_j^{j'} \beta_l^{l'} C_{ijkl} U_{k,l'} T_{,j'} + (J^{-1} \beta_j^{j'} \beta_l^{l'} C_{ijkl} U_k T_{,l'})_{,j'} = 0. \qquad (7)$$

Form invariance of (6, 7) compared to (4a, 4b) leads to the following transformation relation.

$$C'_{i'j'k'l'} = J^{-1} \beta_j^{j'} \beta_l^{l'} C_{ijkl}, \qquad (8a)$$

$$\rho'_{i'j'} = J^{-1} \rho_{ij}, \qquad (8b)$$

$$U'_{i'} = U_i. \qquad (8c)$$

It is interesting to note that these transformation relations are identical with those given by Brun et al. [9], the elastic tensor looses its minor symmetry. Although asymmetric elastic tensor is possible by using high order continuum theory [15], however additional governing equation for the couple stress must be provided. Here, we will solve wave problem within classical elastodynamic theory, namely, Navier's equation, the elastic tensor must be symmetric. Based on equation (8a-c), a symmetric elastic tensor should have the following form

$$C'_{i'j'k'l'} = J^{-1}\beta_i^{i'}\beta_j^{j'}\beta_k^{k'}\beta_l^{l'}C_{ijkl}. \tag{9a}$$

In order to keep the form of (5a) under a general mapping, the transformation relations of the mass density and displacement now become accordingly as

$$\rho'_{i'j'} = J^{-1}\beta_i^{i'}\beta_j^{j'}\rho_{ij}, \tag{9b}$$

$$U'_{i'} = \beta_i^{i'}U_i. \tag{9c}$$

Unfortunately, the transformation relations (9) can not make (5b) be form invariant, therefore they control exactly the path of elastic ray, but only approximately control the energy distribution along the ray. Inserting (9) into (5b), after some algebra, we obtain

$$C_{ijkl}(U_k)_{,l}T_{,j} + (C_{ijkl}U_kT_{,l})_{,j} + \beta_{i'}^m(\beta_i^{i'})_{,j}C_{ijkl}U_kT_{,l} = 0. \tag{10}$$

Compared to (5b), an additional term related to the mapping and its derivative appears, this extra term vanishes only with linear mapping ($(\beta_i^{i'})_{,j} = 0$). If we use (9) to approximately control energy distribution along ray by mapping technique, the error will be smaller for smoother mapping and higher frequency.

To illustrate the theory, elastic ray path and energy distribution must be evaluated, it becomes more involved compared to classical elastic ray theory due to the anisotropic mass density. Ray path can be still calculated with classical ray method [13], to this end, (5a) can be expressed as

$$\Gamma_{sk}U_k = 0, \tag{11}$$

where $\Gamma_{sk} = \rho_{is}^{-1}C_{ijkl}p_lp_j - \delta_{sk}$ is called Christoffel matrix, while $\rho_{is}^{-1}$ and $p_i = T_{,i}$ are the inverse of the mass density matrix and the component of slowness vector, respectively. Based on (11), the path of an incident ray in a smoothly inhomogeneous anisotropic media can be calculated from the eigenvalues $G_i$ and eigenvectors $\vec{g}_i$ of the Christoffel matrix. After determining the initial position and slowness vector of an incident ray, the ray path can be obtained by the following relations:

$$\frac{dx_i}{dT} = \frac{1}{2}\frac{\partial G_m}{\partial p_i}, \tag{12a}$$

$$\frac{dp_i}{dT} = -\frac{1}{2}\frac{\partial G_m}{\partial x_i}, \tag{12b}$$

where $m = 1, 2, 3$ refer qS1 qS2 and qP waves, respectively. The travel time $T$ is chosen to be the curve parameter along the ray.

To evaluate the energy transfer along ray, (5b) is multiplied by $g_i$ and rearranged into

$$2V_i A_{,i} + A(V_i)_{,i} = 0, \tag{13}$$

where $V_i = C_{ijkl} p_l g_k g_j$, and $A$ is the scalar, complex-valued amplitude function, it relates displacement to eigenvector by

$$U_i = A g_i. \tag{14}$$

Equation (13) is called transport equation for an inhomogeneous medium. In classical ray tracing method [13], (13) is evaluated by introducing an elementary ray tube with a specified ray and a family of rays nearby, as shown in Fig. 1. Under a coordinate transformation, a vector $a_i(x_i)$ is transformed into $a'_{i'}(x'_{i'}) = \beta_i^{i'} a_i(x'_{i'})$. Similarly a ray vector $t_i(x_i)$ in a virtual space will be transformed into $t'_{i'}(x'_{i'}) = \beta_i^{i'} t_i(x'_{i'})$ in a physical space. On the other hand, we have

$$\begin{aligned} V'_{i'} &= J^{-1} \beta_i^{i'} \beta_j^{j'} \beta_k^{k'} \beta_l^{l'} C_{ijkl} \beta_{l'}^{r} T_{,r} \beta_{k'}^{s} g_s \beta_{j'}^{t} g_t \\ &= J^{-1} \beta_i^{i'} C_{ijkl} p_l g_k g_j \\ &= J^{-1} \beta_i^{i'} V_i . \end{aligned} \tag{15}$$

Since in a virtual space, $V_i$ is the mass-density-rescaled group velocity vector and has the same direction as $t_i$ [13], so $V'_{i'}$ in the physical space will follow the same direction as $t'_{i'}$ after transformation. In this context, after expressing (13) in ray coordinates, we can apply divergence theorem on $(V_i)_{,i}$ in the elementary ray tube, therefore the same process as classical elastic ray method can then be followed. The amplitudes along the ray can be obtained by using [13]

$$A(T_{n+1}) = \left[ \frac{\Omega^{\perp}(T_n)}{\Omega^{\perp}(T_{n+1})} \right]^{\frac{1}{2}} A(T_n), \tag{16}$$

where $\Omega^{\perp}$ is the cross-sectional area of the elementary ray tube in a 3-D case, as shown in Fig. 1, while in a 2-D case, $\Omega^{\perp}$ is degenerated into the length of one line element connecting two points situated on two close rays. More detailed explanation can be found in the reference [13].

In the following, a 2-D elastic rotator is designed by using the transformation relation (9). The mapping for the rotator is proposed as

$$\begin{aligned} r' &= r, \\ \theta' &= \theta + f(r)\theta_0, \end{aligned} \tag{17}$$

where $\theta_0$ is a rotation angle and $f(r)$ is a five-order polynomial with $f(a)=1$, $f(b)=f'(a)=f'(b)=f''(a)=f''(b)=0$, where $a$ and $b$ are the radius of the inner and outer boundaries of the rotator. This choice can make the boundaries impedance matched. In the computation, $a=0.1\,m, b=0.35\,m$. The background is an isotropic elastic material with normalized Lame constants $\lambda=2.3, \mu=1$ and the density $\rho=1$. A harmonic S wave is emitted from a line source with a small circle of radius $r=0.01$m. Using (12) and (16), the ray paths and amplitudes are calculated by Matlab software. 80 rays of the S-wave with a wave length $l=0.015\,m$ emitted from the source are traced, the results are as shown in Fig.2. A full wave simulation is also performed, the calculated ray path (not shown in the figure) is shown to be identical as that predicted by our transformation ray theory, as expected. The total displacements $\left|\sqrt{U_1^2+U_2^2}\right|$ (representing energy) on four specific rays with the incidence angles $0°, -22.5°, -45°$ and $-67.5°$ calculated by using transformation ray method (16) (labeled by ray method) and full wave simulation (labeled by full wave) are given in Fig.3. The results show that although some errors occur in the rotator, overall the amplitudes calculated by these two methods agree well. This example confirms our theoretical founding: the transformation relation (9) can be used to control exactly the ray path, also approximately well to control the energy distribution along the ray.

In summary, we have proposed a transformation ray theory to control high frequency elastic waves; the transformed elastic stiffness is symmetric. With the proposed transformation relations, the ray path can be controlled exactly in a desired way, the amplitude along rays can also be controlled in an approximate but acceptable way. Finally, it is interesting to find that the transformation relation (9) is identical to that in the reference [12] with a different method. Application of the proposed method can be anticipated for example to seismic protection and health monitoring technique where high frequency approximation is valid.

**Acknowledgments:** this work was supported by the National Natural Science Foundation of China (10832002, 11172037), and the National Basic Research Program of China (2006CB601204).

**Figures:**

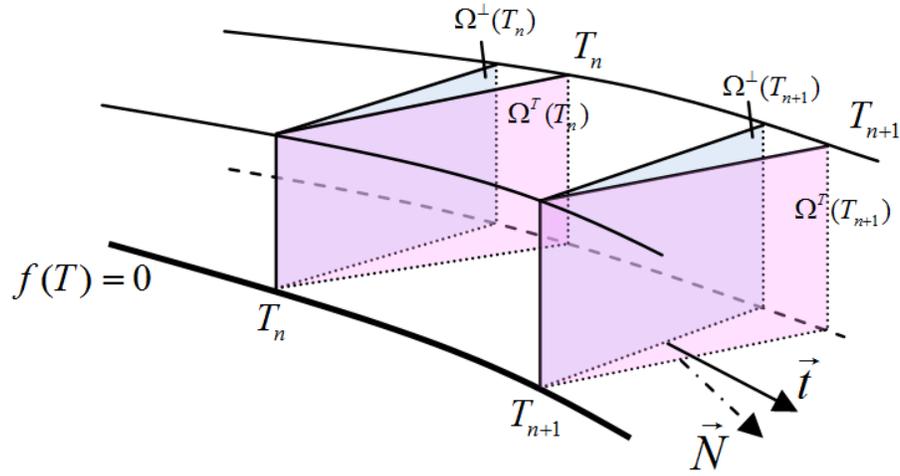

Fig1. A sketch of an elementary ray tube in inhomogeneous anisotropic medium with an elastic ray $f(T)=0$ and a family of rays nearby. $T_n$ and $T_{n+1}$ are two points on the ray used in the recurrent computation. $\Omega^T$ denotes a cut from the ray tube by the wave fronts and $\Omega^\perp$ is the cross section of the ray, which can be calculated through some geometry relation by using $\Omega^T$. $\vec{t}$ is the unit vector perpendicular to $\Omega^\perp$, representing the direction of the ray.

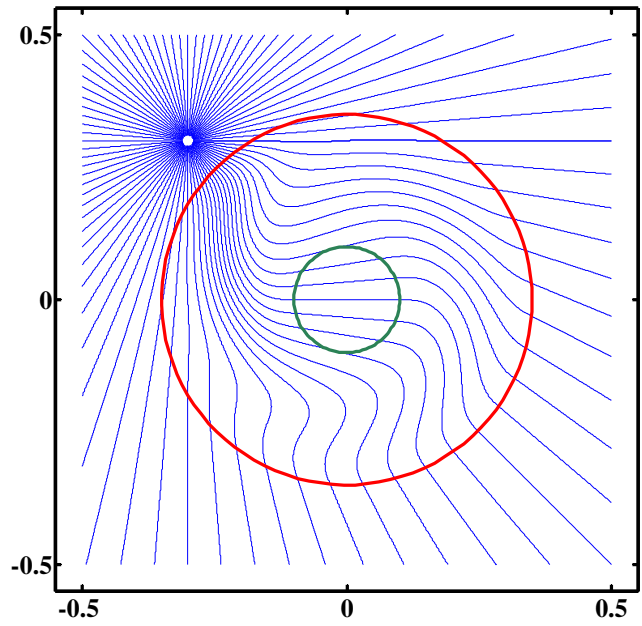

Fig2. Ray traces of the elastic rotator calculated by transformation ray method, they are same as those evaluated by full wave simulation.

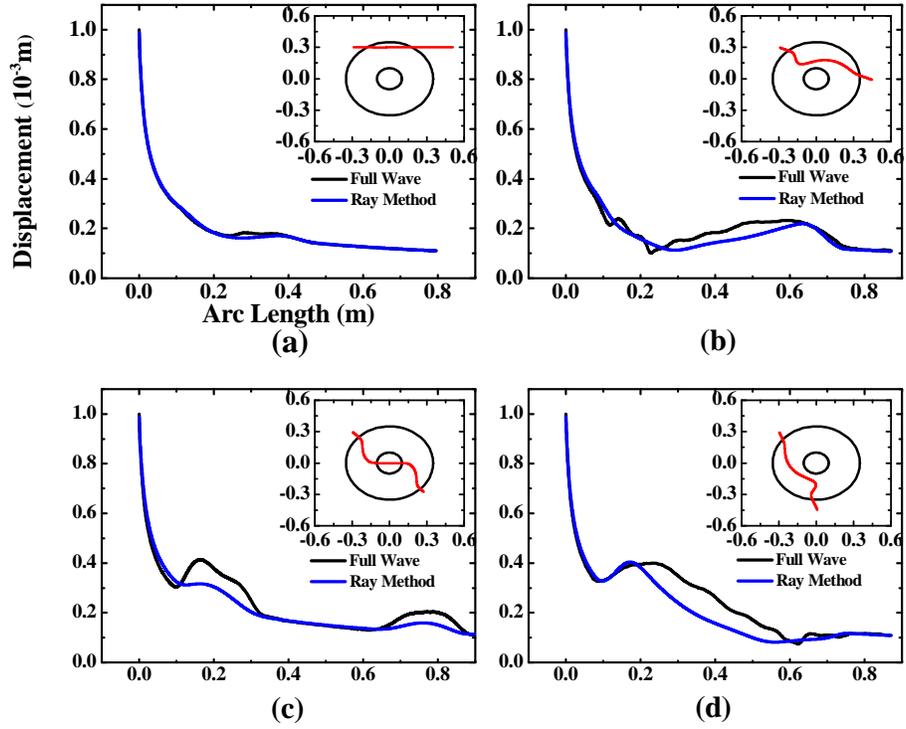

Fig3. Total displacement $\left|\sqrt{U_1^2 + U_2^2}\right|$ along the rays evaluated by transformation ray theory (blue), full wave simulation (black) at incident angles: $0°$ (a), $-22.5°$ (b), $-45°$ (c) and $-67.5°$ (d).